\documentclass{aa}
\usepackage{graphicx}
\def\msun{{\,\rm M_\odot}}

\def\lambdabar{\protect\@lambdabar}
\def\chem#1#2{$\rm{}^{#1}\kern-0.8pt#2$}
\def\reac#1#2#3#4#5#6{$\rm\,{}^{#1}\kern-0.8pt{#2}\,({#3}\,,{#4})\,
{}^{#5}\kern-0.8pt{#6}\,$}
\def\gsimeq{\,\,\raise0.14em\hbox{$>$}\kern-0.76em\lower0.28em\hbox
{$\sim$}\,\,}
\def\lsimeq{\,\,\raise0.14em\hbox{$<$}\kern-0.76em\lower0.28em\hbox
{$\sim$}\,\,}
\def\be{\begin{equation}} 
\def\ee{\end{equation}}
\def\beqy{\begin{eqnarray}}
\def\eeqy{\end{eqnarray}}
\def\bmlet{\begin{mathletters}}
\def\emlet{\end{mathletters}}

\def\A&A#1#2#3{ {\it Astron. Astrophys.} {\bf #2}, #3 (#1)}

\begin{document}

%
   \title{The production of short-lived radionuclides by new non-rotating and rotating
Wolf-Rayet model stars}

   \author{M. Arnould
           \inst{1},
           S. Goriely
           \inst{1},
           \and
           G. Meynet
           \inst{2} 
          }

   \offprints{M. Arnould}
   \institute{Institut d'Astronomie et d'Astrophysique, Universit\'e
              Libre de Bruxelles, CP 226, B-1050 Brussels, Belgium.
              \and
              Observatoire de Gen\`eve, CH-1290 Sauverny, Switzerland
                     }

   \date{Received ---; accepted ---}

\abstract{It has been speculated that WR winds may have
contaminated the forming solar system, in particular with short-lived radionuclides (half-lives in
the approximate $10^5 - 10^8$ y range) that are responsible for a class of isotopic
anomalies found in some meteoritic materials.} {We revisit the
capability of the WR winds to eject these radionuclides using new models of single non-exploding WR stars with metallicity $Z = 0.02$.}{ The earlier
predictions for non-rotating WR stars are updated, and models for rotating such
stars are used for the first time in this context.}{ We find that (1) rotation has no significant influence on the short-lived radionuclide production  by neutron capture during the core He-burning phase, and (2) \chem{26}{Al}, \chem{36}{Cl}, \chem{41}{Ca}, and \chem{107}{Pd} can be  wind-ejected by a variety of WR stars at relative levels that are compatible with the meteoritic analyses for a period of free decay of around $10^5$  y between production and incorporation into the forming solar system solid bodies.} {We confirm the previously published conclusions that the winds of WR stars have a radionuclide composition that can meet the necessary condition for them to be a possible contaminating agent of the forming solar system.  Still, it remains  to be demonstrated from detailed models that this is a sufficient condition for these winds to have provided a level of pollution that  is compatible with the observations.} 
 
 \titlerunning{WR production of short-lived radionuclides}   
\authorrunning{M. Arnould et al.}

\maketitle

%

\section{Introduction}
\label{intro}
 
The decay of short-lived and now extinct radionuclides with half-lives from about $10^5$ to
approximately $10^8$ y has left identifiable traces in various meteoritic materials. They take the  form of excesses of the daughter nucleus, leading to isotopic anomalies with respect
to the bulk solar system composition. 

Some of these anomalies are hosted by solids of solar system origin.  They are generally interpreted
in terms of the injection of thermonuclearly produced now extinct radionuclides in a live form into the solar
nebula, followed by their being trapped in condensing solids. In these views, important
information can be gained about some exciting astrophysical questions related to the
formation and early history of the solar system, and in particular about the time
$\Delta^\ast$ elapsed between the last astrophysical event(s) able to affect the
composition of the solar nebula and the solidification of some of its material. 

A competing model calls for spallative production of these short-lived
radionuclides by the interaction of energetic particles with gas and/or dust in the
protosolar cloud and/or in the solar nebula itself. Of course, in this framework, no
chronology can be derived for nucleosynthetic events just before the formation of solar system solids.  A dominant spallative
origin of most extinct radionuclides is not presently favoured. A noticeable exception is provided by \chem{10}{B}, whose production is only due to spallation. In fact, the solar system irradiation is claimed by Chaussidon et al. (2006) on the grounds of the analysis of Li (in the search for signatures of \chem{7}{Be} in situ decay) in an Allende CAI inclusion. The spallation model will not be considered further here.

The reader is referred to Goswami \& Vanhala (2000) and Goswami et al. (2005) for reviews of the thermonuclear and spallative models. From these, it clearly appears that all scenarios show successes and face failures in the reproduction of the observational data.

The information gained from the
study of solar system solids has been complemented with the discovery that a variety of
short-lived radionuclides have been trapped in grains of supposedly circumstellar origin
and have decayed in-situ before incorporation into meteorites, leaving isotopic
anomalies in the daughter element (e.g. Bernatowitz \& Zinner 1997, Savina et al. 2004).

Supernovae (of Type II), Asymptotic Giant Branch, and Wolf-Rayet (WR) stars have
been envisioned as possible nucleosynthetic progenitors of the short-lived
radionuclides. All these scenarios rely on a variety of speculative arguments. The  WR model of concern in this paper certainly does not escape this state of affairs. Very qualitative arguments have been presented by Arnould et al. (1997; Paper I) and Arnould et al. (1997a; Paper II) in support of the plausibility of one or several WR stars
(possibly in OB associations) being able to inject in a live form into the forming solar system at least some of the  short-lived radionuclides that are extinct by now. 
These WR stars could have been the concomitant triggers for the formation of the Sun and of other
(low-mass) stars. This triggering in an OB association has the
interesting property of avoiding the fortuitous event that is called for in some scenarios
and, in particular, in the picture of the AGB pollution of the solar system. 

The plausibility of the WR scenario has received some further support from speculations that have appeared after Papers I and II concerning the possibility of the solar system having formed in an OB association. In particular,  from a detailed study of the Orion OB1 association, Walter et al. (2000) state  that {\it `planetary systems like our own may be
most likely to form in environments like OB associations'}. They claim
that {\it `there is circumstantial evidence that our Sun may have formed in an OB
association 4.6 Gyr ago'}. This very same conclusion is reached by Hollenbach et al.
(2000) who consider as {\it `quite possible that the solar nebula could have spent its
early life in a stellar cluster, where a nearby O or B star may have photoevaporated the
outer nebula in timescales $t \lsimeq 10^7$ years'}. If indeed solar system-type structures could form in an OB association, the WR contamination scenario still requires that such stars can exist in these associations. This has been discussed in some detail by Kn\"odlseder et
al. (2002). Besides the clear fact that many such associations contain more or
less large amounts of O-type stars that are massive enough to evolve to the WR phase,
some of them are seen to contain some WR stars of the WC subtype of more direct interest
in the question of the radionuclide production (Papers I and II; see also
Sects.~\ref{wrnuclides} and \ref{th_obs}). 

It has to be emphasized that the WR contamination may very well result from only one such star if it was close enough to the solar system. Needless to say, the probability of contamination at the required level conceivably increases with the number of suitable WR stars. The WR star number in an OB association is in fact  time-dependent, as some O-star members can transform into WR stars in the course of their evolution. As an
example, two WC stars are now seen in the OB association Cyg OB1. As stressed by Kn\"odlseder et al.  (2002), this population could
increase over periods of some My commensurable with the time it takes for
a massive star to evolve from the O- to the WR-phase. Let us also recall that WR stars are known to be copious grain producers, as reviewed by e.g. Arnould et al. (1987a) (see also Marchenko et al. 2002, 2003; Monnier et al. 2002). It is not known at this time if at least some of such highly isotopically anomalous grains have found their way into meteorites.
 
In view of the above, it appears interesting to revisit the question of the production of the short-lived radionuclides discussed in Papers I and II on the grounds of new models for non-rotating WR stars in their non-explosive wind phase, which ends in practice at the core He exhaustion, the later evolutionary phases being too short for a steady mass loss to be significant. We also extend the discussion by calculating for the first time the s-process yields from rotating massive stars going through the WR phase. We want to make clear that a large enough production of radionuclides is a {\it necessary} condition for WR stars to be seen as plausible contaminating agents of the early solar system, but is by no means a sufficient condition for an efficient enough pollution. 
 
Section~\ref{wrmodels} briefly summarises the main characteristics of the new WR models,
with special emphasis on the rotating models. Section~\ref{wrnuclides} presents the new
predictions for the WR radionuclide yields, and Sect.~\ref{th_obs} compares the predictions of the
non-rotating and rotating models with observations. Conclusions are drawn in Sect.~\ref{conclusions}. 

\section{The new non-rotating and rotating WR models in brief}
\label{wrmodels}

The main observational and theoretical characteristics of the WR models of interest here
have already been discussed and reviewed in Papers I and II. We just
concentrate on the main aspects of new models, and their main differences
with the former ones. For a detailed comparison between the different models, the reader is referred to Meynet \& Maeder (2003, 2005). Special emphasis is put on the inclusion of rotation in some of
the model stars. 

The physical ingredients of the present models are the same
as those used by Meynet \& Maeder (2003).
The main difference between the present models and the non-rotating ones used in Papers I and II
 concerns the mass loss rates during the WR phase, which have been decreased by a factor of 2 to 3.
 This reduction relates to the consideration of clumping  in the wind (Nugis \& Lamers 2000; Hamann \& Koesterke 1999). The mass loss rates for O--type stars have been revised 
(and in general reduced) as well (Vink et al. 2000, 2001).

\begin{figure}
\centerline{ \includegraphics[scale=0.42]{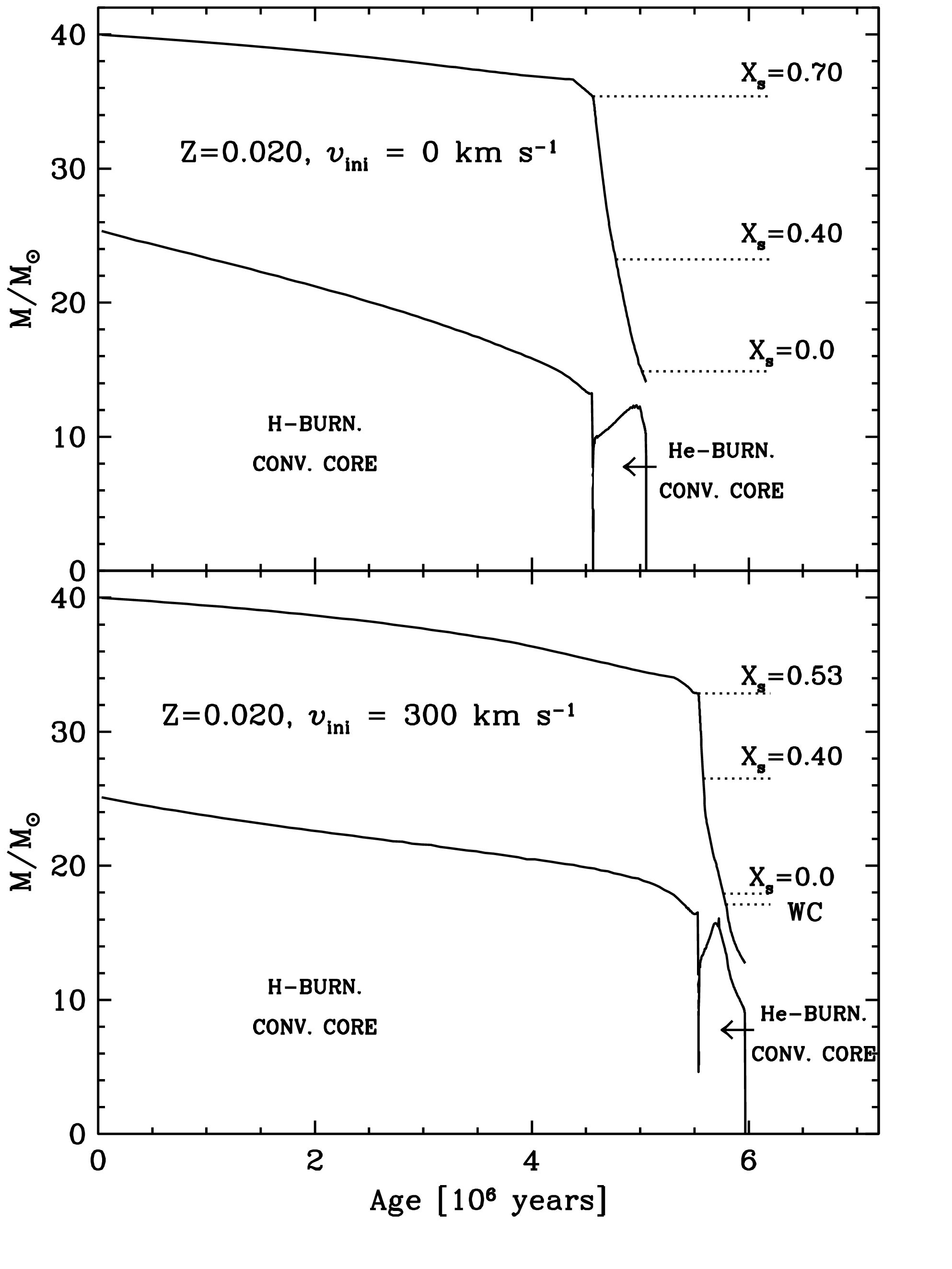}}
\caption[]{Evolution of the total mass $M_{\rm TOT}$ (upper solid lines of each panel) and of the convective core mass
$M_{cc}$ (lower solid lines) for a non-rotating ($v_{\rm ini} = 0$) and for a rotating ($v_{\rm ini} = 300$ kms$^{-1}$)star  with initial mass $M_{\rm i} =
40 \msun$ and metallicity $Z = 0.02$, as predicted by the new evolutionary models. The indicated values of the surface hydrogen mass fraction $X_{\rm s}$ correspond from top to bottom of each panel  to the end of the main sequence, to the start of the WR phase, and to the H surface exhaustion. The label WC identifies the entry point of the rotating model into the WC phase. The non-rotating model does not enter this stage }
\label{structure}
\end{figure}

Another important difference with Papers I and II concerns the inclusion of rotation in some of our star models. Since the effects of rotation on the evolution of massive stars
has already been discussed in detail by Meynet \& Maeder (2003; 2005), we briefly recall the main
points of relevance here. Figure~\ref{structure} illustrates the effects of rotation on the structural evolution of a  $M_{\rm i} = 40 \msun$  with metallicity $Z = 0.02$ from the zero age main sequence (ZAMS) to the end of the core He-burning phase. For this star, as well as for all the other models considered here, a  ZAMS rotational velocity  $v_{\rm ini} = 300$ kms$^{-1}$ is adopted. This initial velocity leads to time-averaged equatorial velocities on the main sequence (MS) that are well within the observed range of 200 to 250 kms$^{-1}$.

As illustrated in Fig.~\ref{structure}, the models with and without rotation exhibit four main differences:
 
 \noindent (1)  The decrease with time in the mass of the convective H-burning core of the rotating model 
is slower than in the non-rotating case. This comes from rotational diffusion, which continuously supplies  hydrogen to the convective core and thus maintains its larger mass. The MS lifetime is consequently longer.

\noindent (2) As a result of (1), the initial He-burning core  is more massive in the rotating model.
 
\noindent  (3) The surface abundances of the non-rotating model at the end of the MS  keep their original values. Changes appear only when the post-MS stellar winds uncover stellar layers contaminated with nuclear burning ashes. The non-rotating star enters the WR phase only when nearly half of its mass has been wind-ejected. With rotation,  the surface composition already changes during the
MS phase, and the star  transforms into a WR, while much less mass has been removed by stellar winds.
This comes from the fact that the typical WR abundance pattern is obtained
in the rotating model not as a result of the removal of the H-rich envelope by stellar winds, 
but mainly as the consequence of rotational mixing. Rotation thus helps the star to enter the WR phase at an earlier evolutionary stage. As a result, the WR lifetime is longer.

\noindent (4) The non-rotating 40 M$_\odot$  star never enters the WC-WO phase characterised by the appearance of  He-burning products at the star surface. This contrasts with the rotating model, which loses  about 4 M$_\odot$ of its material during the WC phase.

In  short,  rotation lowers the critical mass for which a star of a given metallicity can enter the WR phase, and lengthens this stage. As discussed by  Meynet \& Maeder (2003) for $Z = 0.02$ metallicity stars, rotation in fact allows the observed characteristics of WR stars to be accounted for, in contrast to non-rotating models with state-of-the-art  mass loss rates. The same conclusion holds at lower and higher  metallicities (Meynet \& Maeder 2005).

\begin{figure}
 \includegraphics[scale=0.30]{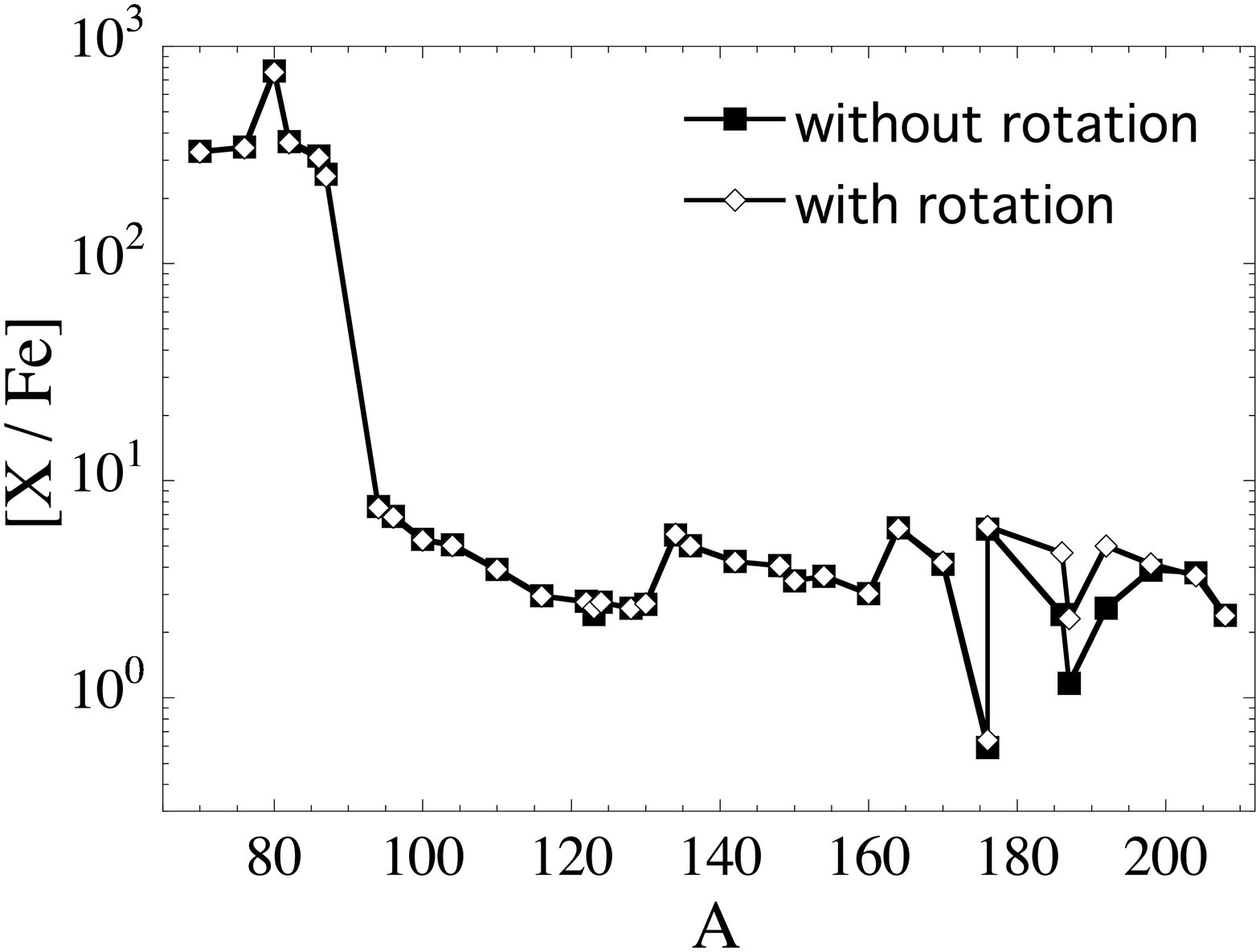}
\caption[]{Mass fractions relative to iron of the (stable) s-nuclides (mass number $A$)
in the convective core at He exhaustion of the non-rotating and rotating $60
\msun$ stars}
\label{sprocess}
\end{figure}

Concerning the production of the trans-iron nuclides of most direct relevance here, Fig.~ \ref{sprocess} shows the s-nuclide abundances at the end of He burning in the
core of a non-rotating and a rotating $60 \msun$ star. They are derived from a classical post-processing nucleosynthesis calculation. Note that diffusion is neglected in the s-process abundance evaluations for the models with rotation, even if this effect is properly taken into account in the stellar model computations. This approximation is required in order to make the evaluation of the abundances of the large number of nuclides implied in the s-process reaction network tractable. 

The present calculations confirm the well-known result that only the s-nuclides up to
the Zr region ($A \lsimeq 90$) are produced in substantial amounts in these massive star
conditions. It also demonstrates that rotation has no impact on the
s-nuclide yields, at least not in the selected range of ZAMS rotational velocities ($0 \leq
v_{\rm i} \leq 300$ kms$^{-1}$). This conclusion
holds for the other stars considered here. From this result, one suspects
that the mass of short-lived radionuclides ejected in the winds of the WR stars is
unaffected by rotation either. Section~\ref{wrnuclides} confirms this inference.   

\section{Results}
\label{wrnuclides}

For all the considered stars that go through the WR phase, we follow the procedure
of Paper I to calculate the wind composition at each
time, and in particular during all the WC and WO evolutionary stages, during which the
radionuclides of interest here are wind ejected. In the following, no distinction is made
between these two stages, generically referred to as the WC-WO phase. Only $Z = 0.02$ stars
are considered here.

The computed stellar evolutionary phases allow us to calculate the composition of the ejecta in a self-consistent way at each time, along with the integrated mass
$M_{\rm R}^{\rm w}(t_1,t)$ of radionuclide R ejected by the wind between instants $t_1$
and $t$ through

\begin{equation}
M_{\rm R}^{\rm w}(t_1,t)=\int\limits_{t_1}^{t} X_{\rm R}^s(t') \exp\left (
-\frac{t - t'}{\tau_{\rm R}}\right )|\dot{M}(t')|dt'. 
\label{eq_wind}
\end{equation}

\noindent In this equation, $X_{\rm R}^s(t)$ is the surface mass fraction of R at time $t$,
and
$|\dot{M}(t)|$ is the instantaneous mass loss rate. Of course, through the stellar
evolution calculations, time $t$ can be univocally translated in terms of the remaining
stellar mass $M(t)$ (Fig.~\ref{structure}). On the other hand, $\tau_{\rm R}$ is the
radionuclide mean life, defined from its half-life $t_{1/2}({\rm R})$ by $\tau({\rm R}) =
t_{1/2}({\rm R})/{\rm {ln}}2$. It has to be stressed that laboratory half-lives apply in
wind conditions. In contrast, $\tau$ may depend more or less drastically on temperature
and density inside  stars (e.g. the review by Arnould \& Takahashi 1999, and references
therein). This dependence is taken into account to calculate the radionuclide
surface content, especially between formation and ejection times; as described below, this
effect is of particular importance for the fate of $^{205}$Pb. Equation~\ref{eq_wind} can be
trivially applied to stable nuclides by setting $\tau_{\rm R} \rightarrow \infty$.

Frequent use is made of times $t_{b\rm ZAMS}$, $t_{b\rm WC}$, and $t_e = t_{e\rm WC}$ at the
start of the ZAMS,  the WC phase, and termination of the WC-WO phase, the last time
corresponding in practice to the pre-supernova stage. In the following, the total wind-ejected mass
$M_{\rm R}^{\rm w}(t_{b\rm ZAMS},t_{e\rm WC})$ of species R is simply denoted as $M_{\rm
R}^{\rm w}$. Of course, all the radionuclides and neighbouring stable species of
interest in this work are not necessarily coeval in the WR ejecta at all evolutionary
stages. More specifically, the H-burning \chem{26}{Al} and \chem{27}{Al} ashes are
ejected at the WN stage (e.g. Meynet et al. 1997) along with the stable s-process
nuclides assumed to be present in the material of the star at its birth. As the latter
cannot be affected by any nuclear processing before the WC stage, it is considered that
their relative yields are in solar proportions. These approximations are considered to
be accurate enough for our purpose. At the onset of the WC-WO phase, the WR wind starts
carrying the s-process radionuclides and stable nuclides produced in the He-burning
core. Some \chem{27}{Al} may also continue to be ejected, in contrast to \chem{26}{Al},
which is very quickly destroyed by (n,p) reactions in the course of central He burning
(e.g. Meynet et al. 1997). 

\begin{figure}
\includegraphics[scale=0.30]{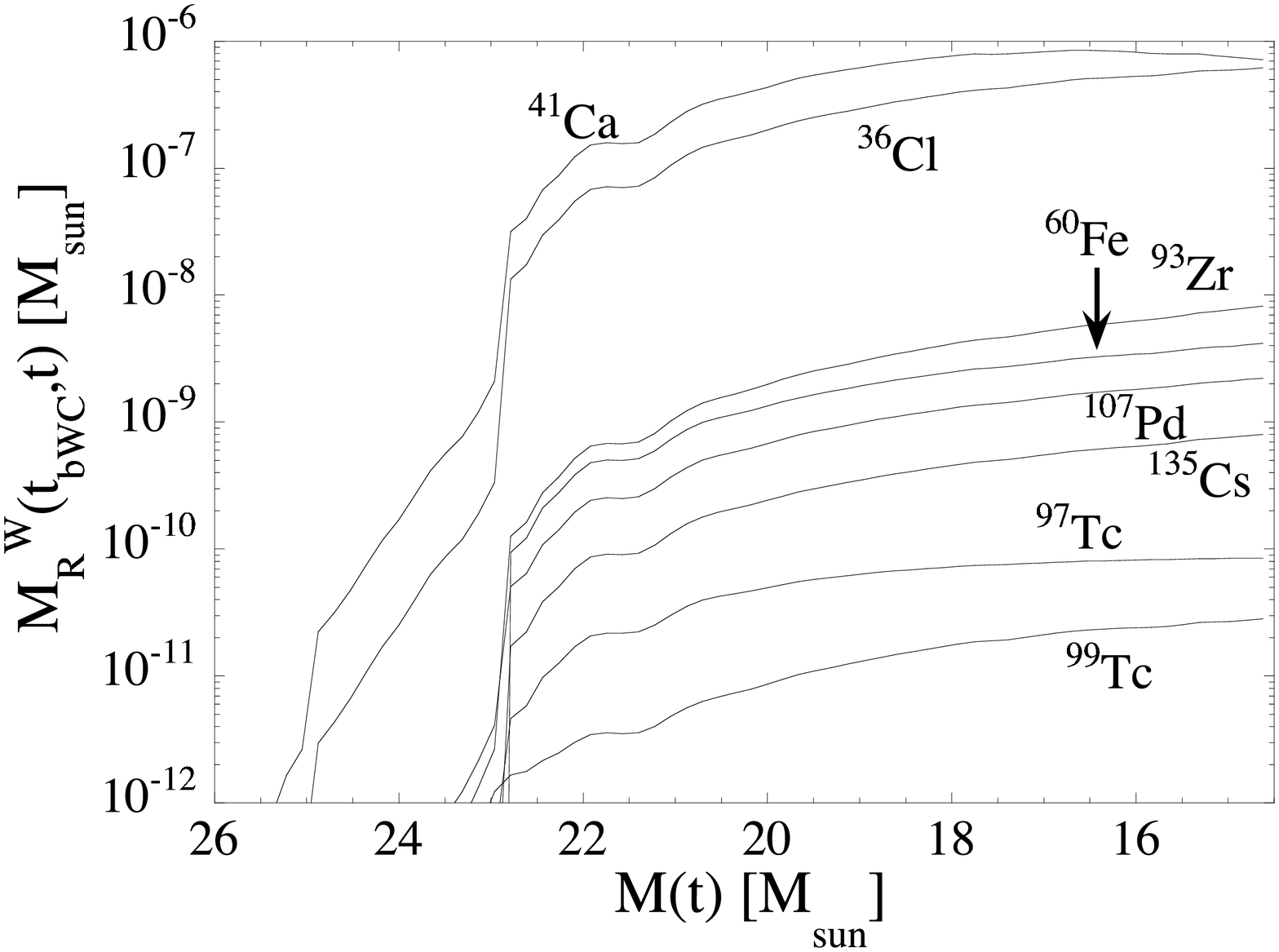}
\caption[]{Values of $M_{\rm R}^{\rm w}(t_{b\rm WC},t)$ for some radionuclides ejected
by the wind of the $Z = 0.02$ rotating $60 \msun$ WR star (Fig.~\ref{structure})
from the start of the WC phase ($t = t_{b\rm WC}$) up to time $t$ (or remaining stellar
mass $M(t)$). For $t \leq t_{b\rm WC}$, the WR wind is free of these radionuclides. }
\label{wind_nuclides}
\end{figure}
 
Figure \ref{wind_nuclides} displays the evolution of $M_{\rm R}^{\rm W}(t_{b\rm WC},t)$
for some short-lived radionuclides produced by the He-burning s-process in the
rotating $60 \msun$ model star. The non-rotating case gives
similar results, and is not displayed. 

The most important difference between Fig.~2 of Paper I based on former WR models and
Fig.~\ref{wind_nuclides} concerns the \chem{205}{Pb} wind content, which is negligible
in our new models.  This is in marked contrast to the conclusions of Paper I.  Some
discussion of this drastic change in our conclusions is clearly in order. Like the other
short-lived radionuclides of interest here,
\chem{205}{Pb} is produced by the s-process in the convective
He-burning core. Unlike the others, \chem{205}{Pb} is in danger of
being destroyed after leaving the convective core and before being ejected from the star
through its wind, i.e. during its residence in the radiative buffer layer between the
convective core and the stellar surface (see Fig.~\ref{structure}). This situation arises
from the very special decay properties of the \chem{205}{Pb} - \chem{205}{Tl} pair (Yokoi
et al. 1985). As illustrated by  Fig.~2 of Mowlavi et al. (1998), the \chem{205}{Pb}
electron capture rate $\lambda_e$ and the \chem{205}{Tl} $\beta$-decay rate $\lambda_-$ are
quite sensitive functions of temperature/density. In absence of neutrons, as it is the case
in the radiative buffer, \chem{205}{Pb} can be prevented from destruction by electron
captures only if it can be replenished by the \chem{205}{Tl} $\beta$-decay, i.e. if
$\lambda_- \geq \lambda_e$. This happens for temperatures ranging from about 0.8 to
$1.5 \times 10^8$ K for electron number densities increasing from $10^{25}$ to $10^{27}$
cm$^{-3}$. This requirement is not fulfilled in the radiative buffers predicted by our
new models. As a result, \chem{205}{Pb} decays there on timescales $1/\lambda_-$ that
are shorter than the calculated radiative buffer lifetimes. The level of survival or
destruction of \chem{205}{Pb} before its eventual ejection in the interstellar medium
is thus drastically dictated by these lifetimes, which in turn is sensitive to
the extent of the convective core and and the mass loss rate, if not to multidimensional
phenomena that could only develop close to the WR surface. As all these complicated
aspects of the WR physics are poorly mastered, we conclude this time in the light of
our new results that it is quite risky to draw strong conclusions on the
\chem{205}{Pb} yields from WR stars.

\section{Comparison with observations}
\label{th_obs}

The short-lived radionuclides with half-lives shorter than $10^8$ y for which there is
strong (\chem{26}{Al}, \chem{41}{Ca}, \chem{53}{Mn}, \chem{60}{Fe}, \chem{107}{Pd},
\chem{182}{Hf}) or suggestive evidence that needs confirmation ({\chem{36}{Cl},
\chem{92}{Nb}, \chem{99}{Tc}, \chem{205}{Pb}) for their presence in live form in the
early solar system have been reviewed and discussed by Goswami \& Vanhala (2000). We compare the initial solar system radionuclide abundances they
provide with our predictions in a form similar to the one already adopted in Papers I
and II.

\subsection{The case of\, \chem{\rm 107}{\rm Pd}}
\label{pd}

Our predicted radionuclide yields are normalized to the ratio
\chem{107}{Pd}/\chem{108}{Pd} = $4.5 \times 10^{-5}$. This is the value at the time of
formation of the Ca-Al-rich refractory inclusions (CAIs) inferred by Goswami \& Vanhala
(2000) on grounds of the `canonical' value of $2 \times 10^{-5}$ used in Papers I and II,
and derived from the analysis of iron meteorites.  Use of the CAIs as reference points relates to the fact that they are expected to be among the first solid phases to have condensed in the
solar system. In the following, we denote by (R/S)$_0$ the ratio of the abundance
of the radionuclide R to a neighbouring stable nuclide S at the time of formation of the
early solids in the solar system. 

\begin{figure}
\includegraphics[scale=0.30]{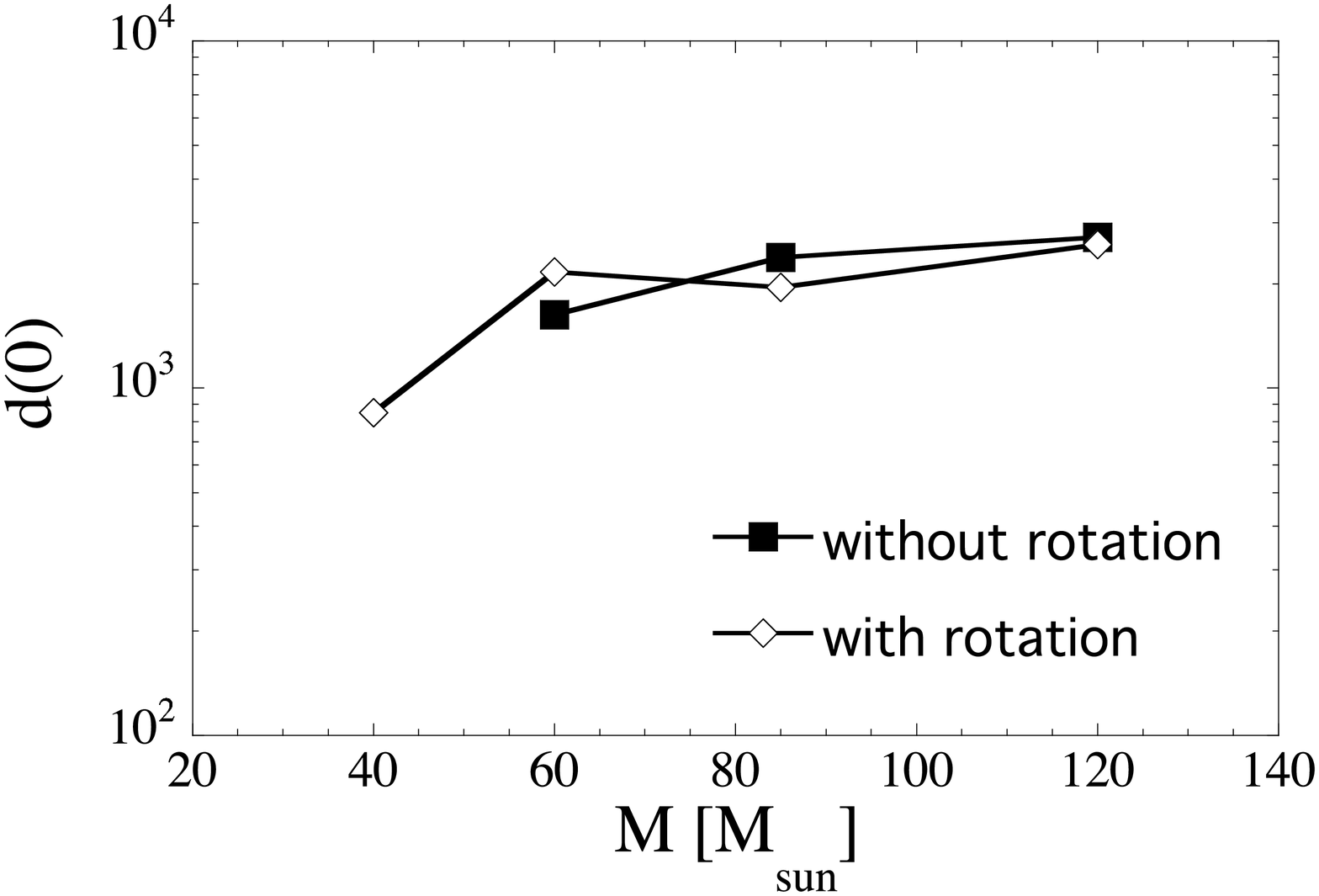}
 \caption[]{Values of $d(0)$ [Eq.~\ref{eq_dilution}] for the rotating and non-rotating
 $Z = 0.02$ stars that are predicted to go through the WC-WO phase. They are derived from
the ratio (\chem{107}{Pd}/\chem{108}{Pd})$_0 = 4.5 \times 10^{-5}$ at the time of formation
of the CAIs (Goswami \& Vanhala 2000)}
\label{fig_dilution}
\end{figure}

On the other hand, it is assumed that the
\chem{107}{Pd} present in the wind of a WR star at the end of its WC-WO phase can find
its way into the CAIs after a period $\Delta^*$ of free decay. In this picture, the
(\chem{107}{Pd}/\chem{108}{Pd})$_0 = 4.5 \times 10^{-5}$ ratio can be accounted for if a
normalization (`dilution') factor $d$ is applied to the value of the
\chem{107}{Pd}/\chem{108}{Pd} calculated in the WR wind. The $d(0) \equiv d(\Delta^* =
0)$ values derived in the absence of a free-decay period are presented in
Fig.~\ref{fig_dilution} for all the considered non-rotating and rotating $Z = 0.02$ 
stars that go through the WC-WO phase, and under the simplifying assumption of a complete
mixing of the wind material ejected between the ZAMS and the end of the computed
evolutionary sequences. Figure~\ref{fig_dilution} shows that $d(0)$ values around
$10^3$  are allowed for. This conclusion is in line with the findings of
Papers I and II. However, they now appear to be more insensitive to the stellar
mass than concluded there. This especially concerns the (non-rotating) $120 \msun$
case. We also find that $d(0)$ does not differ greatly between the non-rotating and
rotating cases, which merely confirms that the s-process efficiency is largely insensitive
to rotation (Fig.~\ref{sprocess}). Note that the dilution factors that have
to be applied for $\Delta^* \neq 0$ can be trivially derived from $d(0)$ following

\begin{equation}
d(\Delta^*) = d(0) \exp(-\Delta^*/\tau_{^{\rm 107}\rm Pd}),
\label{eq_dilution}
\end{equation}

\noindent where $\tau_{^{\rm 107}\rm Pd}$ is the \chem{107}{Pd} lifetime.

As discussed in Paper I, the assumption of thorough mixing of all the
material lost by stellar wind, which is made in order to obtain the dilution factors of
Fig.~\ref{fig_dilution}, might be replaced by a scenario in which the complete mixing is
limited to the WC-WO phase. In such an extreme case, which may have some astrophysical
relevance (see Paper I), larger dilution factors would be obtained, the pre-WC
\chem{107}{Pd}-free wind being suppressed.

From the dilution factors just discussed, we attempted in Paper I to estimate
if indeed there is any chance of contamination of the
protosolar nebula by isotopically anomalous WR wind material, and in particular by
\chem{107}{Pd}. It was concluded from qualitative considerations about this highly complex
question that such a possibility is not utterly farfetched. This contamination might
even be concomitant to the triggering of the Sun and solar system formation. In fact, as
already stressed in Sect.~\ref{intro}, some recent observations even lend support
to these views.  

This scenario may also be tested more quantitatively by numerical simulations that are
reviewed by Goswamy \& Vanhala (2000). In particular, one has to address the questions of
(1) the formation of the solar system triggered by shock waves, radiation, or wind from
nearby (massive) stars interacting with interstellar globules or with molecular clouds,
(2) the efficiency of injecting radioactivities into the forming solar system, and
(3) the timescales required for such an injection. This is of course of special
importance when dealing with unstable nuclides. The existing
simulations only provide some hints, as they are not at a sufficient stage of
sophistication and reliability to give definite answers to the above questions. Future investigations will hopefully allow a better appraisal of the merits of the WR scenario compared to the other ones proposed so far for the contamination of the early solar system by short-lived radionuclides.

As in Papers I and II, the dilution factors deduced from the \chem{107}{Pd}
yields are applied to the other short-lived radionuclides of interest here, thus
neglecting possible fractionation effects. 

\begin{figure}
\includegraphics[scale=0.30]{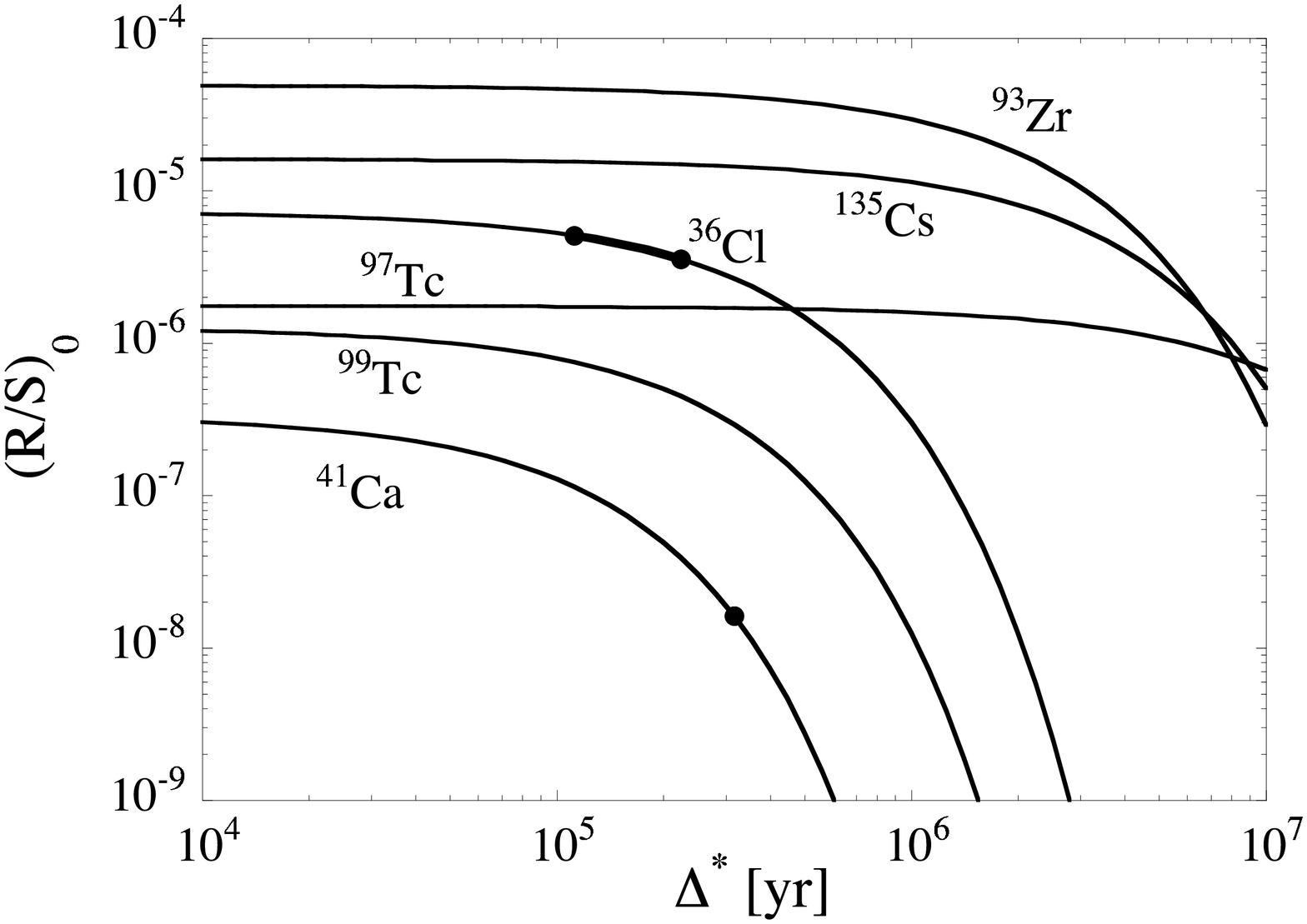}
\caption{Abundance ratios (R/S)$_0$ of various radionuclides R relative
to stable neighbours S versus $\Delta^\ast$ (see main text) for the rotating
$40 \msun$ model star. The curves labelled  \chem{36}{Cl}, \chem{41}{Ca},
\chem{93}{Zr}, \chem{97}{Tc}, \chem{99}{Tc}, and \chem{135}{Cs} refer to the following
(R/S)$_0$ ratios:  \chem{36}{Cl}/\chem{37}{Cl},
\chem{41}{Ca}/\chem{40}{Ca}, \chem{93}{Zr}/\chem{92}{Zr}, \chem{97}{Tc}/\chem{100}{Ru},
\chem{99}{Tc}/\chem{100}{Ru} and \chem{135}{Cs}/\chem{133}{Cs}. All the displayed ratios
are normalized to (\chem{107}{Pd}/\chem{108}{Pd})$_0 = 4.5\times10^{-5}$
(Sect.~\ref{pd}) by applying of a common dilution factor $d(\Delta^\ast)$, the
values of which can be derived from $d(0) = 850$ displayed in Fig.~\ref{fig_dilution} and from Eq.~\ref{eq_dilution}. The black dot and thick solid line correspond to the normalized observed ratios adopted from Fig.~2 of Goswami \& Vanhala (2000) for Ca and from Murty et al. (1997) for Cl}
\label{fig_radio_40}
\end{figure}

\begin{figure}
\includegraphics[scale=0.30]{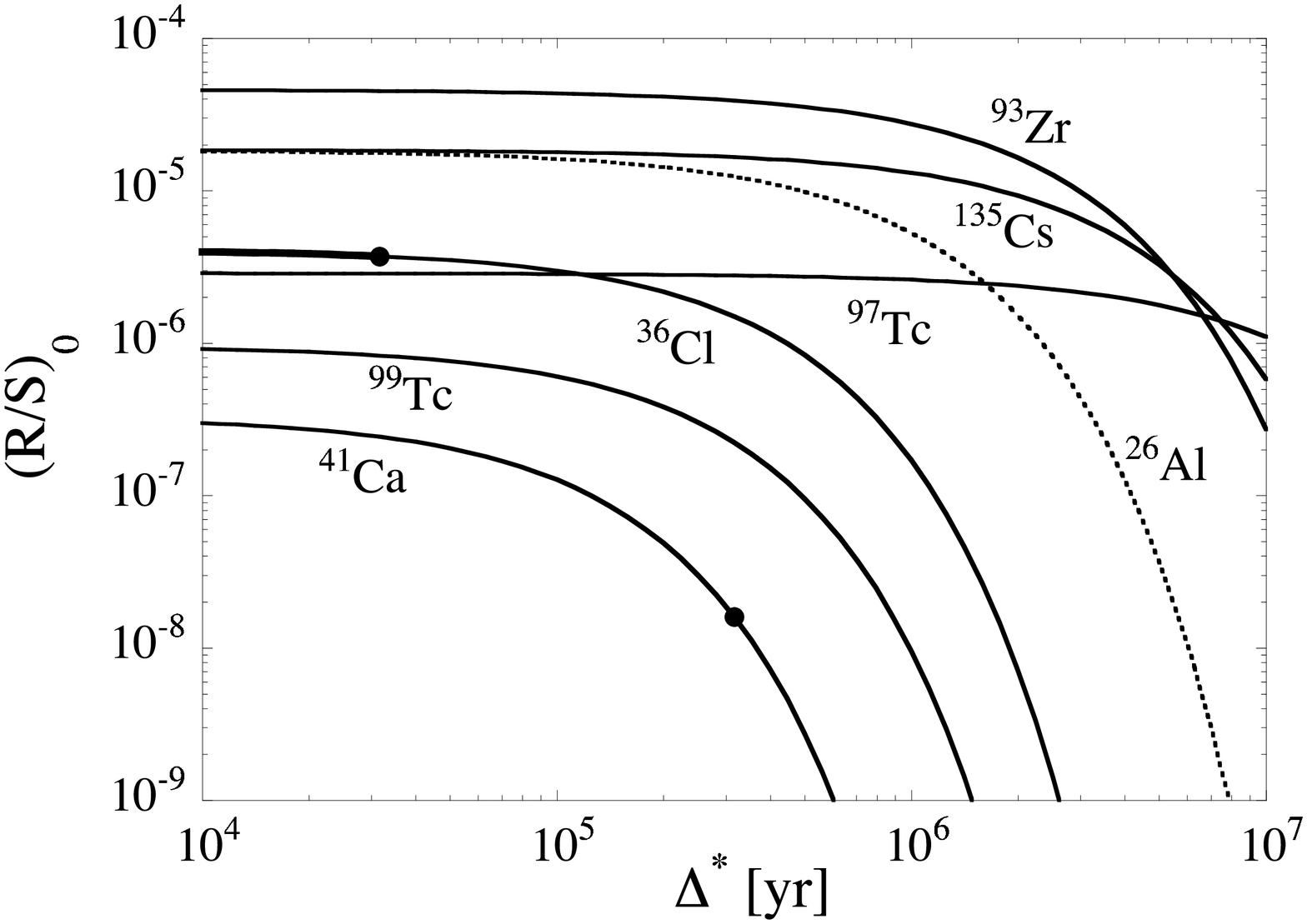}
\caption{same as Fig.~\ref{fig_radio_40}, but for the rotating $60 \msun$ star. In this
case, $d(0) = 2160$. The dashed line corresponds to the \chem{26}{Al}/\chem{27}{Al} ratio calculated by Palacios et al. (2005)}
\label{fig_radio_60}
\end{figure}

\begin{figure}
\includegraphics[scale=0.30]{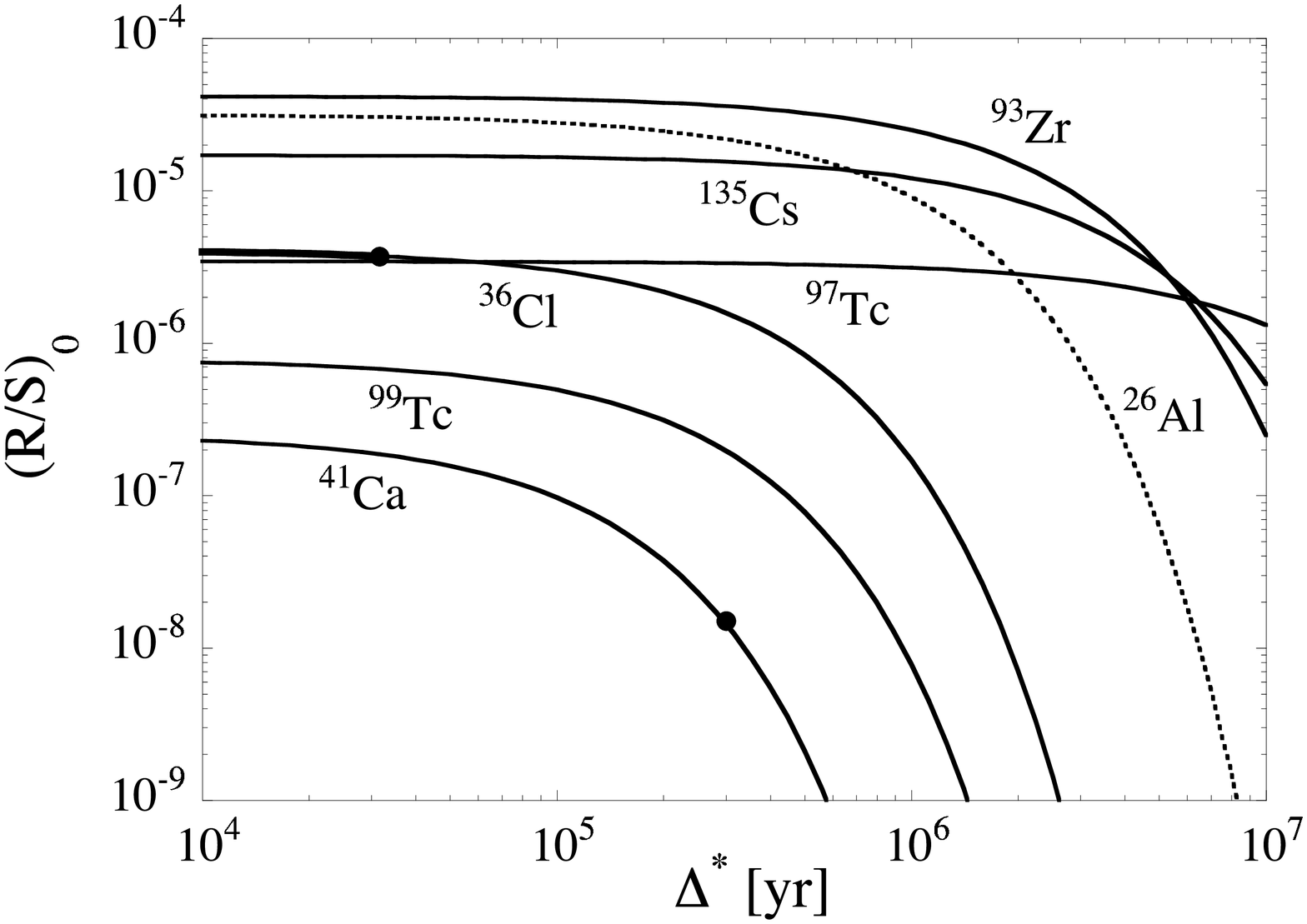}
\caption{same as Fig.~\ref{fig_radio_60}, but for the rotating $85 \msun$ star. In this
case, $d(0) = 1950$}
\label{fig_radio_85}
\end{figure}

\begin{figure}
\includegraphics[scale=0.30]{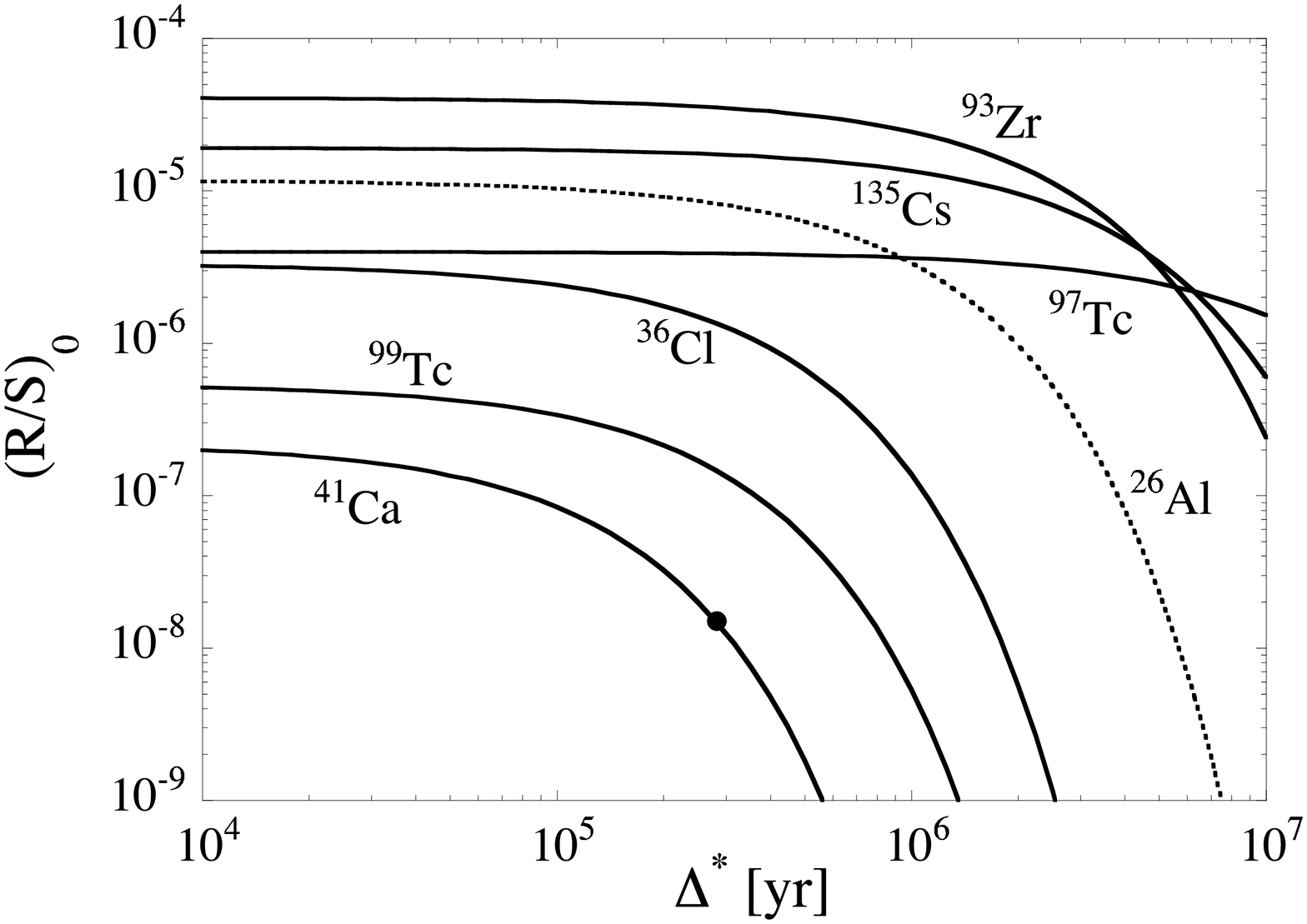}
\caption{same as Fig.~\ref{fig_radio_60}, but for the rotating $120 \msun$ star. In this
case, $d(0) = 2590$}
\label{fig_radio_120}
\end{figure}

\subsection{The case of\, \chem{\rm 26}{\rm Al}}
\label{al}

In contrast to the other short-lived radionuclides considered in this work, \chem{26}{Al} 
of special importance for cosmochemistry, as well as for $\gamma$-ray astrophysics, is not
produced during the core He-burning phase, but is destroyed instead in this environment
by (n,p) reactions (see above). In fact, \chem{26}{Al} is produced in the
H-burning core mainly by \reac{25}{Mg}{p}{\gamma}{26}{Al}, and wind ejected during the WN
evolutionary stage preceding the WC-WO phase (e.g. Meynet et al. 1997).

In the absence of rotation, we calculate the \chem{26}{Al} yields as in Papers I and II.  We confirm the former results that the non-rotating  $Z = 0.02$ stars with $M_{\rm i} \geq 60 \msun$ do not allow the canonical
value (\chem{26}{Al}/\chem{27}{Al})$_0 = 5 \times 10^{-5}$ to be reached. The new $40 \msun$ star never enters the WC-WO phase (see Sect.~\ref{wrmodels}), and consequently does not eject any \chem{26}{Al}. The Paper I $Z = 0.02$ $40 \msun$ star experiences the WC-WO stage instead, but its \chem{26}{Al} yield is insufficient to account for the observed ratio. 

 For the rotating cases,  the recent predictions of Palacios et al. (2005) are adopted, as their Al abundances are self-consistently derived from the stellar models with the proper inclusion of  diffusion effects. Their yields normalized to (\chem{107}{Pd}/\chem{108}{Pd})$_0$ are displayed in Figs.~\ref{fig_radio_60} - \ref{fig_radio_120} . Note that we are not able to draw conclusions for the rotating $40 \msun$ star, due to numerical problems encountered in the calculation of its evolution. The ability of this star to produce enough \chem{26}{Al} thus remains to be investigated.  In Paper I, it was concluded that the \chem{26}{Al} yields increase with metallicity. This question has not been re-examined here. The situation for rotating $Z > 0.02$ models also remains to be scrutinized. 

It has to be recalled that the conclusions above are drawn when the same dilution factors are applied to Al and Pd. This might not be fully
adequate from a physico-chemical point of view. This approximation might also break down
because, as stressed above, \chem{26}{Al} loads the wind while the massive stars are in
their WN stage preceding the WC phase, during which all the other radionuclides
considered here are ejected.

\subsection{The other short-lived radionuclides}
\label{other_radionuclides}

The information on the level of production of the other short-lived radionuclides of
cosmochemical interest that could be ejected along with \chem{107}{Pd} by the WC-WO wind
of the considered rotating stars is summarised in Figs.~\ref{fig_radio_40} -
\ref{fig_radio_120}. These figures display
versus $\Delta^*$ the ratios (R/S)$_0$ of the abundance of the radionuclide R to the one
of a neighbouring stable nuclide at the start of condensation in the solar system. In
all cases, they are normalized to (\chem{107}{Pd}/\chem{108}{Pd})$_0$ (Sect.~\ref{pd}),
the adopted dilution factors $d(\Delta^*)$ at each time being derived from
Fig.~\ref{fig_dilution} and Eq.~\ref{eq_dilution}. We conclude that

\noindent (1) the WC-WO wind of all the considered stars can carry enough \chem{41}{Ca}
($t_{1/2} = 10^5$ y)  to account for (\chem{41}{Ca}/\chem{40}{Ca})$_0 =  (1.38 \pm 0.13)  \times 10^{-8}$
(Goswami \& Vanhala 2000) if $\Delta^\ast \approx 2 - 3 \times 10^5$ y, a value remarkably
independent of $M_{\rm i}$.

\noindent (2) the value
(\chem{36}{Cl}/\chem{37}{Cl})$_0 = (4.4\pm 0.6) \times 10^{-6}$ derived from the 
(\chem{36}{Cl}/\chem{35}{Cl})$_0 = (1.4\pm 0.2)  \times 10^{-6}$ ratio reported (but which still
needs confirmation) by Murty et al. (1997) is obtained for the  $40 \msun$ star if $\Delta^\ast \approx 1 - 2
\times 10^5$ y, close to the value obtained for \chem{41}{Ca}. The observed ratio is reached only for lower  $\Delta^*$ values in the 60 and 85 $\msun$ cases, and is only approached closely in this $\Delta^*$ range for the $120 \msun$ star. 

\noindent (3) in contrast, by far too little \chem{60}{Fe} is synthesized to account
for (\chem{60}{Fe}/\chem{56}{Fe})$_0 \approx 2 \times 10^{-8}$ at the time of CAIs
formation derived by Goswami \& Vanhala (2000).

\noindent (4) more or less large amounts of  \chem{93}{Zr}, \chem{97}{Tc},
\chem{99}{Tc} and \chem{135}{Cs} can also be produced in several cases, but
these predictions cannot be tested due to the lack of relevant
or reliable observations. Note that (i) the \chem{93}{Zr} decay to \chem{93}{Nb} cannot
lead to observable isotopic anomalies in view of the monoisotopic nature of \chem{93}{Nb},
and (ii) the production by the s-process of the neutron-deficient \chem{97}{Tc} may look
surprising. An explanation of this situation can be found in Paper I.

\noindent (5) as discussed in Sect.~\ref{wrnuclides}, no significant amount of
\chem{205}{Pb} is predicted to be ejected by the new non-rotating or rotating WR model
stars.

\section{Conclusions}
\label{conclusions}

This paper presents the first calculations of the s-process in rotating $Z = 0.02$ stars
that are massive enough to become WR stars. It also revisits the predictions of papers I and II for non-rotating models. Here are our main conclusions.

\noindent (1) Rotation has no significant influence on the s-processing during central He
burning. This concerns, in particular, the production of the radionuclides with half-lives
ranging from about $10^5$ to $10^8$ y, whose decay might have left identifiable
traces in meteorites.

\noindent (2) A series of radionuclides (\chem{26}{Al}, \chem{36}{Cl},
\chem{41}{Ca}, \chem{107}{Pd}) can be wind-ejected by a variety of 
WR stars at {\it relative} levels compatible with the meteoritic
observations for a convergent period of free decay $\Delta^\ast$  of the order of
$10^5$ to $3 \times 10^5$ y between production and
incorporation into the first solid bodies forming in the solar system. This result is in line with those of Paper I.

\noindent (3) In contrast, the new (rotating or non-rotating) WR models do not predict any
significant production of \chem{205}{Pb}. This situation is drastically different from the
one described in Paper I. The analysis of the reasons for such a discrepancy leads us to
consider that it is almost impossible at this time to draw strong conclusions about the
true amount of \chem{205}{Pb} possibly ejected by the winds of WR stars. This
unsatisfactory state of affairs relates in particular to the uncertainties in the
description of the He-burning convective cores and in the evaluation of the mass loss rates
during the WR stage.

As in Paper I, the present conclusions are derived without
taking into account the possible contribution from the material ejected by the eventual
supernova explosion of the WR stars being considered. This supernova
might add its share of radionuclides that are not produced
abundantly enough during the WR wind phase. This concerns in particular
\chem{53}{Mn}, \chem{60}{Fe} (see e.g. Limongi \& Chieffi 2006), or \chem{146}{Sm}. One also has to acknowledge
that the above conclusions sweep the possible role of
binarity in the WR yields completely under the rug. Its impact on the predicted \chem{26}{Al}
production and the additional level of uncertainty it generates have been
explored by Langer et al. (1995). 

The WR relative production of radionuclides in quantites that are compatible with
observation for a  convergent value of $\Delta^\ast$ is of course only a necessary,
but not a sufficient, condition for making the contamination of the forming
solar system by the WR winds plausible. We complement this result with the contention, which is
based on qualitative considerations developed in Paper I,  that astrophysically plausible
situations might be encountered such that radionuclides ejected by WR stars could
contaminate the solar system at an absolute level compatible with the observations.
Of course, the interaction of WR stars with their surroundings is extremely complex, and
these complications have to be dealt with in detail before the qualitative
conclusion can be confirmed or rejected that WR stars could indeed be viable ``last minute'' solar system contaminators
 . It is gratifying that this inference seems to receive more
and more support from observation, as recalled in Sect.~\ref{intro}.

\vskip0.5truecm

\noindent {\it Acknowledgements.} S.G. is FNRS Research Associate.


\begin{thebibliography}{}

\bibitem[]{} Arnould, M., \& Takahashi, K 1999, Rep. Prog. Phys., 62, 393
\bibitem[]{} Arnould, M., Paulus, G., \& Meynet, G. 1997, A\&A, 321, 452 (Paper I)
\bibitem[]{} Arnould, M.,Meynet, G., \& Paulus, G. 1997a, Astrophysical
Implications of the Laboratory Study of Presolar Materials, ed. T.J.Bernatowicz, \& E.K.
Zinner E, CP402 (New York: AIP), 179 (Paper II)
\bibitem[]{} Bernatowitz, T.J., \& Zinner, E.K. (eds.) 1997, Astrophysical
Implications of the Laboratory Study of Presolar Materials, CP402 (New York: AIP) 
\bibitem[]{} Chaussidon, M., Robert, F., \& McKeegan, K.D., 2006, Geochim. Cosmochim. Acta, 70, 224
\bibitem[]{} Goswami, J.N., \& Vanhala, H.A.T. 2000,  Protostars and
Planets, ed. V. Mannings, A.P. Boss, \& S.S. Russell (Tucson: Univ. Arizona Press), 963
\bibitem[]{} Goswami, J.N., Marhas, K.K., Chaussidon, M., et al., 2005, Chondrites and the Protoplanetary Disk, ASP Conf. Ser., vol. 341, ed. A.N. Krot, E.R.D. Scott, \& B. Reipurth (San Francisco: ASP), p. 485
\bibitem[]{} Hamann, W.R., \& Koesterke, L. 1999, IAU Coll. 169, eds.  B. Wolf, O. Stahl, \& A.W. Fullerton,
(Berlin: Springer). Also Lecture Notes in Physics, 523, 1999, p. 239
\bibitem[]{} Hollenbach, D.J., Yorke, H.W., \& Johnstone, D. 2000, Protostars and Planets,
ed. V. Mannings, A.P. Boss, \& S.S. Russell (Tucson: Univ. ArizonaPress), 401
\bibitem[]{} Kn\"odlseder, J., Cervi\~no, M., Le Duigou, J.-M., et al. 2002, A\&A,
390, 945
\bibitem[]{} Langer, N., Braun, H., \& Fliegner, J. 1995, Astrophys. Space Sci., 224, 275
\bibitem[]{} Limongi, M., \& Chieffi, A. 2006, New Astron. Rev., in press [arXiv:astro-ph/0512598]
\bibitem[]{} Marchenko, S.V., Moffat, A.F.J., Vacca, W.D., et al. 2002, ApJ, 565,
L59
\bibitem[]{} Marchenko, S.V., Moffat, A.F.J., Ballereau, D., et al. 2003, ApJ,
596, 1295
\bibitem[]{} Meynet, G., Maeder, A. 2003, A\&A, 404, 975
\bibitem[]{} Meynet, G., Maeder, A. 2005, A\&A, 429, 581
\bibitem[]{} Meynet, G., Arnould, M., Prantzos, N., \& Paulus, G. 1997, A\&A, 320,
460
\bibitem[]{} Monnier, J.D., Tuthill, P.G., \& Danchi, W.C. 2002, ApJ, 567, L137
\bibitem[]{} Mowlavi, N., Goriely, S., Arnould, M. 1998, A\&A, 330, 206
\bibitem[]{} Murty, S.V.S., Goswami, J.N., \& Shukolyukov, Yu.A. 1997, ApJ, 475, L65
\bibitem[]{} Nugis, T., \& Lamers, H.J.G.L.M. 2000, A\&A, 360, 227
\bibitem[]{} Palacios, A., Meynet, G., Vuissoz, C., et al. 2005, A\&A, 429, 613
\bibitem[]{} Savina, M.R., Davis, A.M., Tripa, C.E., et al. 2004, Science, 303, 649
\bibitem[]{} Vink, J.S., de Koter, A., \& Lamers, H.J.G.L.M. 2000, A\&A, 362, 295
\bibitem[]{} Vink, J.S., de Koter, A., \& Lamers, H.J.G.L.M. 2001, A\&A, 369, 574
\bibitem[]{} Walter, F.M., Alcal\'a, J.M., Neuh\"auser, R., et al. 2000,  
Protostars and Planets, ed. V. Mannings, A.P. Boss, \& S.S. Russell (Tucson: Univ. Arizona
Press), 273
\bibitem[]{} Yokoi, K., Takahashi, K., \& Arnould, M. 1985, A\&A, 145, 339

\end{thebibliography}
\end{document}